# Unsupervised Learning Based Focal Stack Camera Depth Estimation

**Zhengyu Huang[1,2*], Weizhi Du[1,2*] and Theodore B. Norris[1,2]**
*1.Center for Ultrafast Optical Science, University of Michigan, Ann Arbor, Michigan, 48109, USA*
*2.University of Michigan, Ann Arbor, Michigan, 48109, USA*
** These authors contributed equally.*
zyhuang@umich.edu

**Abstract:** We propose an unsupervised deep learning based method to estimate depth from focal stack camera images. On the NYU-v2 dataset, our method achieves much better depth estimation accuracy compared to single-image based methods.

## 1. Introduction

Camera based depth estimation is an important topic due to applications in augmented reality, target tracking and autonomous driving. In 2017, Srinivasan et al. [1] proposed to use a defocused image as the supervision and trained a depth estimation network end-to-end to predict the scene depths. In 2019, Gur et al. [2] proposed to use a defocused focal stack, instead of single image, as the supervision to train the network. However, both methods use a single image as the input to train the neural networks, which limits their depth estimation performance due to the lack of sufficient depth clues. Recently, a transparent graphene based focal stack camera system was proposed by [3], which can capture focal stack images and three-dimensional information in a single exposure. In this report we propose an unsupervised method to estimate depth using focal stack camera. Compared to single-image based depth estimation, focal stack camera has significantly better depth estimation accuracy.

## 2. Method and Experiments

Fig. 1 illustrates the pipeline of the proposed unsupervised depth estimation method. The input focal stack images are passed into a convolutional neural network (CNN) to estimate a depth map. Then a differentiable focal stack rendering module takes in the estimated depth map and an estimated all-in-focus image (estimated using a Laplacian operator) as inputs to reconstruct the focal stack. Finally, a photometric reconstruction error is used as the network loss and the gradient can be back-propagated to train the CNN.

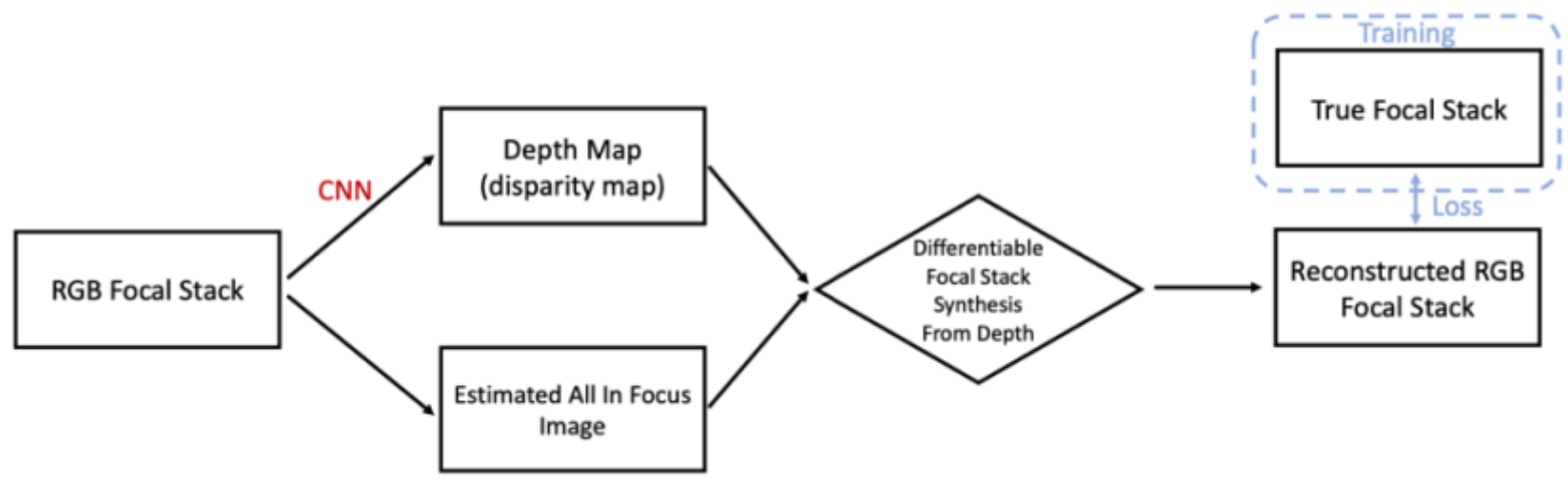

Figure 1. Flow chart of the proposed unsupervised depth from focus method.

We train and evaluate the performance of our proposed method using NYU-v2 dataset [4], which contains 120k RGB images of indoor scenes in depth range [0.7m, 10m]. We used 654 images from a subset of 1449 aligned RGB-depth pairs for testing. Since the NYU-v2 dataset only contains sharp all-in-focus images, we synthesized focal stacks images using these all-in-focus images and ground truth depth. Each focal stack contains 6 focal stack images with their focus distances set to be 0.8 m, 1 m, 1.2 m, 1.6 m, 2.4 m, 5 m. This particular focus distance setting ensures the depth of field of neighboring focal stack images are in contact with each other, but with no overlap. We trained the network for 170k iterations using a batch size of 2, a learning rate of $2 \times 10^{-5}$ using Adam optimizer [5].

## 3. Results

Fig. 2 shows depth estimation results on the test samples using the proposed method. Interestingly, even without direct depth supervision during training, our proposed method can still estimate the depth with good quality. Table 1 compares the depth estimation performance of the single-image based method (row 1) [2] and proposed focal stack based method (row 2). It shows that our proposed method achieves much lower RMSE (root-mean-squared-error) and better $\delta$ depth accuracy [6], demonstrating the advantage of using focal stack for 3D perception purpose. We additionally evaluated our proposed method's performance using ground-truth all-in-focus images (row 3), instead of those estimated from the Laplacian operator. It shows that a better all-in-focus image estimation could further improve the depth accuracy, which is left as a future work.

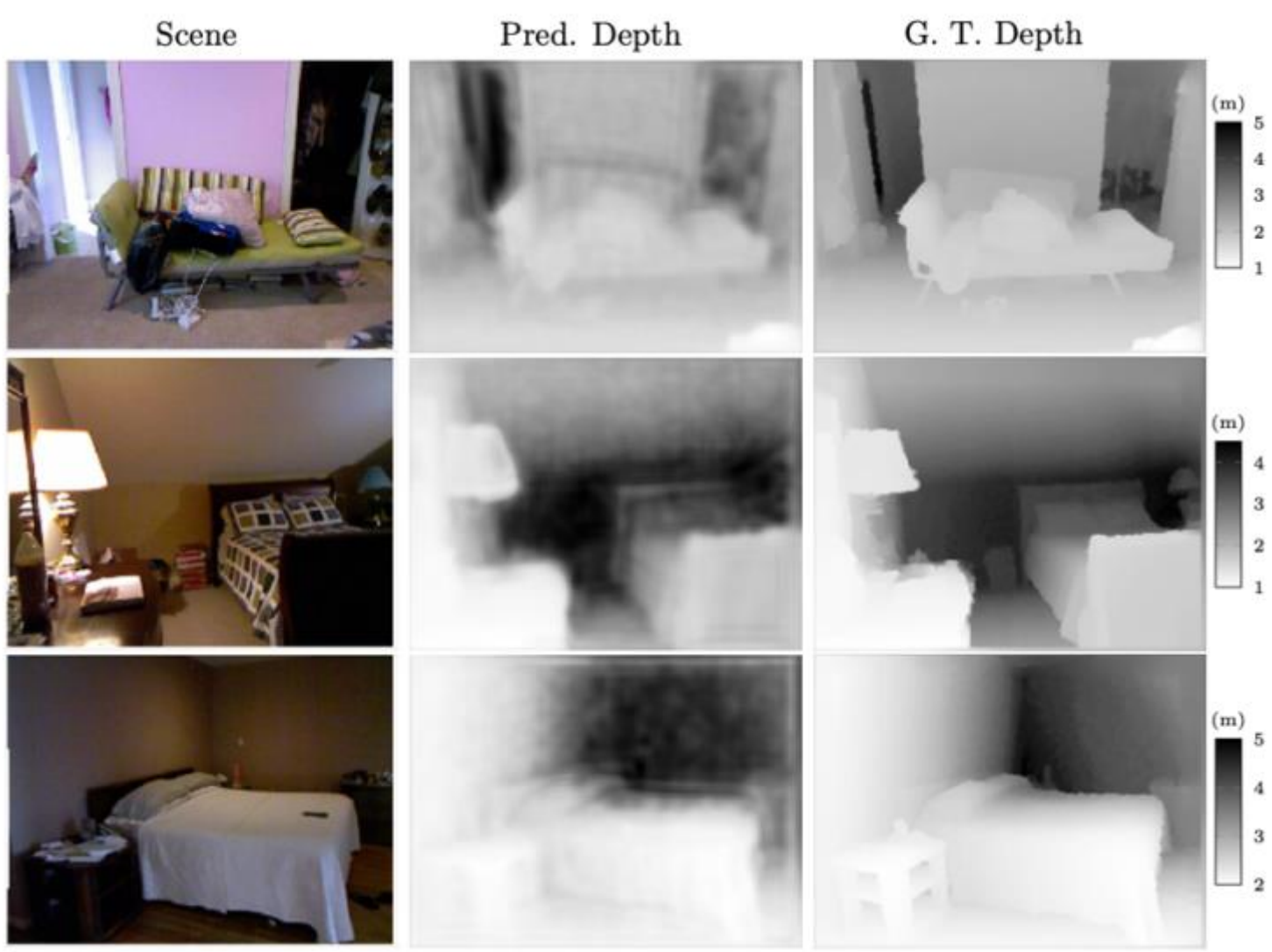

Figure 2. Visualization of the depth estimation result. column 1: test scenes; column 2: predicted depth; column 3: ground truth depth.

| | RMSE (m) | $\delta < 1.25$ | $\delta < 1.25^2$ | $\delta < 1.25^3$ |
|---|---|---|---|---|
| Depth from single image [2] | 0.546 | 0.797 | 0.951 | 0.987 |
| Focal stack with syn. $I_{\mathrm{AIF}}$ | 0.310 | 0.959 | 0.990 | 0.997 |
| Focal stack with g.t. $I_{\mathrm{AIF}}$ | 0.244 | 0.955 | 0.985 | 0.997 |

Table 1. Result of unsupervised depth from focal stack images.

## 3. Conclusion

In this work, we proposed an unsupervised method to estimate a depth map using focal stack camera images. By using a differentiable focal stack synthesis module and a focal stack reconstruction loss, the network can be trained end to end without depth supervision. Numerical experiments show that the proposed method achieves good depth estimation accuracy and outperforms the single-image based unsupervised depth estimation method.

Funding. W. M. Keck Foundation.